\documentclass[epj,nopacs]{svjour}

\usepackage{latexsym}
\usepackage[numbers,sort&compress]{natbib}
\usepackage{graphics}
\usepackage{graphicx}
\usepackage{subfigure}
\usepackage{hyperref}
\usepackage[tbtags]{amsmath}
\usepackage{color}
\usepackage{amsmath}
\usepackage{stfloats}
\usepackage{epstopdf}

\begin{document}

\title{The distinctions between $\Lambda$CDM and $f(T)$ gravity according Noether symmetry}
\author{Han Dong \inst{1} \thanks{\email{donghan@mail.nankai.edu.cn}},
Jiaxin Wang \inst{1} \thanks{\email{jxw@mail.nankai.edu.cn}}
\and Xinhe Meng\inst{1,2} \thanks{\email{xhm@nankai.edu.cn}}
}

\institute{Department of physics, Nankai University, Tianjin 300071, China
\and Kavli Institute of Theoretical Physics China, CAS, Beijing 100190, China}

\date{Received: date / Revised version: date}

\abstract{Noether's theory offers us a useful tool to research the conserved quantities and symmetries of the modified gravity theories, among which the $f(T)$ theory, a generally modified teleparallel gravity,
has been proposed to account for the
dark energy phenomena. By the Noether symmetry approach,
we investigate the power-law, exponential and polynomial forms of $f(T)$ theories.
All forms of $f(T)$ concerned in this work possess the time translational symmetry, which is related with energy condition or Hamilton constraint. In addition, we find out that the performances of the power-law and exponential forms are not pleasing.
It is rational adding a linear term $T$ to $T^n$ as the most efficient amendment to resemble the teleparallel gravity or General Relativity on small scales, ie., the scale of the solar system. The corresponding Noether symmetry indicates that only time translational symmetry remains. Through numerically calculations and observational data-sets constraining, the optimal form $\alpha T + \beta T^{-1}$ is obtained, whose cosmological solution resembles the standard $\Lambda$CDM best with lightly reduced cosmic age which can be alleviated by introducing another $T^m$ term. More important is that we find the significant differences between $\Lambda$CDM and $f(T)$ gravity. The $\Lambda$CDM model has also two additional symmetries and corresponding positive conserved quantities, except the two negative conserved quantities.
}

\maketitle

\section{Introduction}

Diverse sets of astrophysical observations, which include WAMP~\cite{WAMP}, SN Ia~\cite{SNe Ia}, X-ray~\cite{X-ray}, LSS~\cite{LSS} and SDSS~\cite{SDSS} clearly indicate that our universe is currently accelerated expanding. These observations also imply that our universe is spatially flat on large scales, and consists of about 68\% dark energy (DE) with effectively negative pressure, 32\% dust matter (cold dark matter plus baryons), and negligible relativistic constituent (including radiation and neutrinos) at present.

In order to explain the cosmic acceleration, many theories have been proposed phenomenologically and physically. Generally speaking, the approaches can be classified into two broad groups: 1) Modification the energy-momentum tensor of Einstein's equation, where the framework of general relativity (GR) is kept unchanged and specific dark energy fields are invoked. The simplest candidate of dark energy is
a tiny positive time-independent cosmological constant $\Lambda$ with equation-of-state (EOS) parameter $\omega \!\!=\!\!-1$.
Although this is highly consistent with the available observational data-sets,
it faces difficulties such as the microphysical origin,
besides, recent observations favor EOS parameter of constant DE lower than -1~\cite{planck16}.
2) Modified gravity theories provide another way, which present natural unification of the early-stage in action and late-stage acceleration due to different roles of gravitational terms relevant at small and at large curvature. Scalar-tensor theories~\cite{scalar-tensor}, $f(R)$ gravity~\cite{f(R),fR_stability_conditions,fR_review,fR_sss},
and others are studied extensively.
Besides which the $f(T)$~\cite{fT0,fT,fT1,fT11}gravity, where $T$ is the so-called torsion scalar, has been proposed recently and attracted much attention. In the case $f(T)\!\!=\!\!T$, $f(T)$ theory can be directly reduced to the Teleparallel Equivalent of General Relativity (TEGR)~\cite{TEGR,WCW} which was first propounded by Einstein in 1928~\cite{einstein} to unify gravity and electromagnetism by introducing a vierbein field along with the concept of absolute parallelism or teleparallelism.

Constraints on $f(T)$ gravity~\cite{fT3} by latest observational data-sets, dynamical behavior~\cite{fT4} and cosmic large scale structure~\cite{fT_LSS}, relativistic neutron star~\cite{fT_relativistic_stars}, matter bounce~\cite{fT_bounce} and perturbations~\cite{pt} in $f(T)$ gravity framework, static spherical symmetry solutions, validity of Birkhoff's theorem~\cite{fT_birkhoff} and the EOS parameter crossing the phantom divide~\cite{fT_w} are investigated by previous researches.

Without fundamental principles, most of the models generated from modified gravity theories in previous work were proposed phenomenologically, which seem arbitrary. Specific solutions of the field equations depend on explicit expressions of MG (modified gravity) theories, which are important for gaining insight into the mathematical and physical contents of the theories. In our present work, we concern about whether there exist some fundamental principles capable of providing reasonable and acceptable forms of modification, special quantities related to the cosmological or astrophysical observations are also expected.

Particularly, the well known Noether's theorem is suitable for the purpose mentioned above, which has been extensively studied with allowed models given by the Noether symmetry approach, ie., in scalar field cosmology~\cite{scalar_Noether}, non-minimally coupled cosmology~\cite{non-minimally_coupled_Noether}, $f(R)$ theory~\cite{fR_Noether}, $f(T)$ theory~\cite{fT_Noether}, scalar-tensor theory~\cite{scalar-tensor_Noether} and quantum cosmology~\cite{quantum_Noether}.

The complete solutions via Noether symmetry~\cite{Noether_symmetry} need to be confirmed, or expand it to General Noether symmetry~\cite{General_Noether_symmetry}. Physical contents of conserved quantities and their relations to cosmological or astrophysical observations are of great interest. Notice that not all of the conserved quantities derived from Noether's are physically meaningful, nor satisfy the astrophysical constraints, and there may also exist degeneracy phenomena between Noether symmetries.

In this article, we investigate in detail several manifestations of the $f(T)$ gravity according to Noether's theory and observational constraints. This piece of work is arranged as follows:
In the next section we will briefly review the $f(T)$ theory. The Noether symmetries for high-order $f(T)$ manifestations are thoroughly researched in section three, ie., the exponential form and polynomial expression. Consistency of those models will be checked with astrophysical data-sets are also provided. In section four we analyze in detail the $\Lambda$CDM model and distinctions between $\Lambda$CDM and $f(T)$ gravity according Noether symmetry. In the last sections, we shall summarize some conclusions with discussions.

\section{brief introductory to $f(T)$}

In the teleparallel gravity, the vierbein field
$\mathbf{e}_{i}(x^{\mu})$ plays the role of dynamical variables,
which are defined as the orthogonal basis of the tangent space
at each point $x^{\mu}$ in the manifold, namely,
$\mathbf{e}_{i}\cdot \mathbf{e}_{j}\!\!=\!\!\eta_{ij}$,
where $\eta_{ij}\!\!=\!\!diag$(1,-1,-1,-1) is the Minkowski metric.
The vierbein vectors can be expanded in space-time coordinate basis:
$\mathbf{e}_{i}\!\!=\!\!e^{\mu}_{i} \partial_{\mu}$,
$\mathbf{e}^i\!\!=\!\!e^i_\mu{\rm d}x^\mu$. According to the convention,
Latin indices and Greek indices, both running from 0 to 3,
label the tangent space coordinates and the space-time
coordinates respectively. The components of vierbein are related by
$e_{\mu}^i e^{\mu}_j\!\!=\!\!\delta^{~i}_{j}$ and
$e_{\mu}^i e^{\nu}_i\!\!=\!\!\delta_{\mu}^{~\nu}$.

The metric tensor is related to the vierbein
field $\mathbf{e}_{i}(x^{\mu})$ as
\begin{equation}\label{basic}
g_{\mu\nu}=\eta_{ij} e_{\mu}^i e_{\nu}^i,
\end{equation}
which can be equivalently expressed as $\eta_{ij}\!\!=\!\!g_{\mu\nu} e^{~\mu}_{i} e^{~\nu}_{j}$.
The definition of the non-vanishing torsion tensor is given then by
\begin{equation}
T^{\rho}_{~\mu\nu}=e_i^{\rho}(\partial_{\mu}e^i_{\nu} - \partial_{\nu}e^i_{\mu}).
\end{equation}
Evidently, $T^{\rho}_{~\mu\nu}$ vanishes in the Riemann geometry
since the Levi-Civita connection is symmetric
with respect to the two covariant indices.
Differing from general relativity,
teleparallel gravity relies on Weitzenb\"ock connection,
which is defined directly from the vierbein.

In order to get the action of the teleparallel gravity, it is
convenient to define the contorsion:
\begin{equation}\label{contorsion}
K^{\mu\nu}_{~~\rho}=-\frac{1}{2}(T^{\mu\nu}_{~~\rho}-T^{\nu\mu}_{~~\rho}-T_{\rho}^{~\mu\nu}).
\end{equation}
Moreover, instead of the Ricci scalar $R$ for the Lagrangian density in general relativity, the torsion scalar $T$ describing the teleparallel Lagrangian density is defined by
\begin{equation}\label{T}
T=S_{\rho}^{~\mu\nu} T^{\rho}_{~\mu\nu},
\end{equation}
where
\begin{equation}\label{S}
S_\rho^{~\mu\nu}=\frac{1}{2}(K^{\mu\nu}_{~~\rho}
+\delta_\rho^{~\mu}T^{\theta\nu}_{~~\theta}
-\delta_\rho^{~\nu}T^{\theta\mu}_{~~\theta}).
\end{equation}
The modified teleparallel action of $f(T)$ gravity is expressed as
\begin{equation}\label{action}
 I=\frac{1}{16\pi G}\int {\rm d}^4x~e\,f(T),
\end{equation}
where $e\!\!=$det$(e^i_{\mu})\!\!=\!\!\sqrt{-g}$.
The teleparallel gravity can be obtained by setting $f(T)\!=\!T$.
This modification is expected possibly to provide a natural way
to explain the astrophysical observations,
especially for the mysterious dark energy, as a motivation.

\section{Noether symmetry for $f(T)$}

The field equations in modified gravity are partial differential equations (PDEs), and solutions of which can be provided and investigated by the Noether's theorem in a systematic and mathematical way, considered as the most elegant and systematic approach to compute conserved quantities, given by Noether in 1918. The conservation laws play a vital role in the study of physical phenomena so far. The Noether theorem states that any differentiable symmetry of the action of a physical system has a corresponding conservation law. The main feature of this theorem is that it may provide meaningful information regarding the conservation laws in the theory, ie., angular momentum can be explained by the rotational symmetry, and energy conservation due to the time translational invariance. Moreover, the Noether symmetries are defined as transformations in the configuration space which preserve the Lagrangian form exactly. In the generalized symmetries, the configuration space is extended to the first or higher order of time derivatives of coordinates, which do not preserve the Lagrangian form exactly, except the modulo contact forms and exact differentials.

Our present work is built on the base of flat FRW space-time,
and the explicit description of torsion scalar $T\!=\! -6 H^2$.
The method of Lagrange multipliers is adopted to set $T$ as a constraint
of the dynamics, we can obtain the point-like FRW Lagrangian
\begin{equation}\label{Lagrangian f(T)}
    L = a^{3}(f-f_{T}T) - 6f_{T}a\dot{a}^2.
\end{equation}
The configuration space is $\{t,a,T\}$, so the Noether symmetry generator reads
\begin{equation}\label{symmetry generator f(T)}
    X = \tau(t,a,T)\frac{\partial}{\partial t} + \psi(t,a,T)\frac{\partial}{\partial a} + \phi(t,a,T)\frac{\partial}{\partial T},
\end{equation}
where $\tau$, $\psi$ and $\phi$ are smooth functions of independent variables $t$ and the canonical coordinates $a$ and $T$.
Then the first prolongation $X^{[1]}$ can be defined as
\begin{equation}\label{first prolongation}
    X^{[1]} = X + \dot{\psi}(t,a,T)\frac{\partial}{\partial a} + \dot{\phi}(t,a,T)\frac{\partial}{\partial T} .
\end{equation}
The Lagrangian quantity should satisfy the following equation:
\begin{equation}\label{Noether equation f(T)}
    X^{[1]} L + (D\;\tau) L = D\; B(t,a,T),
\end{equation}
where $B(t,a,T)$ is the gauge function, and $D$ indicates the total derivative.
The first integral is known as the conserved quantity
associated with $X$ by time integral, which is defined as
\begin{equation}\label{conserved quantity f(T)}
    I = B - \tau L - (\psi - \tau \dot{a})\frac{\partial L}{\partial \dot{a}} - (\phi - \tau \dot{T})\frac{\partial L}{\partial \dot{T}}.
\end{equation}
Expanding Eq.~(\ref{Noether equation f(T)}) with the Lagrangian of $f(T)$, a determined system of linear PDEs can be obtained:
\begin{eqnarray} \label{PDEs f(T)}
  a f_{T}\tau_{a} &=& 0, \nonumber \\
  a f_{T}\tau_{T} &=& 0, \nonumber \\
  a f_{T}\psi_{T} &=& 0, \nonumber \\
  2 a f_{T} \psi_{a} + f_{T} \psi + a f_{TT} \phi - a f_{T}\tau_{t} &=& 0, \nonumber \\
  -12 a f_{T}\psi_{t} + a^3(f-f_{T}T)\tau_{a} &=& B_{a}, \nonumber \\
  a^3(f-f_{T}T)\tau_{t} &=& B_{T}, \nonumber \\
  3a^2(f-f_{T}T)\psi - a^3 f_{TT}T \phi + a^3(f-f_{T}T)\tau_{t} &=& B_{t}.
\end{eqnarray}
In the following we will investigate several specific forms (models) of manifestation of $f(T)$ theory with corresponding Noether symmetries, cosmological solutions and astrophysical data-sets constraints on relevant parameters of each model.

\subsection{Case \uppercase\expandafter{\romannumeral1}: $f(T)\sim T^n$}
\subsubsection{model depiction}

In this case where we assume that $f(T)\sim T^{n}$, which is the simplest and commonly considered expression adopted in the $f(T)$ theory~\cite{fT,fT3}. According to the introduction above, taking the form to the determined system of linear PDEs~(\ref{PDEs f(T)}), the solutions of $\tau$, $\phi$, $\psi$ and $B$ read
\begin{eqnarray}
  \tau &=& C_{1}t + C_{2}, \\
  \psi &=& -\frac{1}{3}a \bigg[n C_{3} a^{-3/2n} + C_{1} (1-2n)\bigg], \\
  \phi &=& \bigg(-2 C_{1}+ C_{3}a^{-3/2n}\bigg)T, \\
  B &=& const,
\end{eqnarray}
where the index number $n$ is assumed to be a real value, rather than an integer, and the $C_{i}$ are the normalized orthogonal coefficients. By setting one of the $C_{i}$ to $1$ and the others to $0$, we can get the corresponding Noether symmetry generators, which are deduced as
\begin{eqnarray}
  X_{1} &=& t\frac{\partial}{\partial t} + \frac{2n-1}{3}a\frac{\partial}{\partial a} - 2T\frac{\partial}{\partial T}, \label{ftc1x1}\\
  X_{2} &=& \frac{\partial}{\partial t}, \label{ftc1x2}\\
  X_{3} &=& -\frac{1}{3} n a^{1-3/2n}\frac{\partial}{\partial a} + a^{-3/2n}T\frac{\partial}{\partial T}, \label{ftc1x3}
\end{eqnarray}
where Eq.~(\ref{ftc1x2}) apparently indicates energy conservation in the universe, while the physical meanings of Eqs.~(\ref{ftc1x1}) and~(\ref{ftc1x3}) are a little more complicated and vague, which may suggest that time invariance being conjugated with cosmic inflation and the change of torsion.

The commutator relations of the set $\{X_1, X_2, X_3\}$ can be presented by the non-vanishing Lie bracket $[X_i, X_j]$,
\begin{equation}
    [ X_1 ,  X_2 ]  = -X_2,\quad  [ X_1 ,  X_3 ]  = \frac{1-2n}{2n} X_3, \quad[ X_2 ,  X_3 ]  = 0 ,
\end{equation}
thus the commutator relations of the set are closed.
Moreover, the corresponding first integrals read
\begin{eqnarray}
  I_{1} &=& a^3 T^n t (2n-1) + 4 a^3 H T^{n-1} n(2n-1), \label{I1} \\
  I_{2} &=& a^3 T^n (2n-1), \label{I2} \\
  I_{3} &=& -4 a^3 n^2 a^{-3/2n} H T^{n-1}, \label{I3}
\end{eqnarray}
among which, $I_{2}$ and $I_{3}$ are time independent integrals.
Setting $a_0\!=\!1$ (normalization of the scale factor at present), $T \!=\!-6 H^2$ and $1 + z =\frac{1}{a}$ (relation between red-shift and the scale factor) conventionally, Eqs.~({\ref{I2}}) and~(\ref{I3}) finally indicate the same cosmological solution, which reads
\begin{equation}
    H = H_{0} (1+z)^{3/2n}, \label{sol1}
\end{equation}
where $H_0$ represents the Hubble parameter at present. When $n$ equals to one, an universe filled with non-relativistic matter only can be retrieved, according to $\Omega_m\!=\!1$ for $H(z)\!=\!H_0\sqrt{\Omega_m (1+z)^3}$. From a much practical view, we can forecast that the observational data-sets constraints must favor $n\!>\!1$ in order to provide explanations for dark energy and relativistic constituents (which include photons and neutrinos).

The third solution (corresponding to symmetric generator $X_{3}$) has been discussed in a previous research~\cite{fT_Noether},
where the configuration space considered was incomplete. So the other two solutions (corresponding to symmetric generators $X_1$ and $X_2$) are not found there. The $I_2$ can be regarded as the quantity of energy conservation according to the time translational symmetry $X_{2}\!=\!\frac{\partial}{\partial t}$.
The corresponding Hamiltonian $E\!=\!I_{2}$ can also be derived by Eq.~(\ref{E}),
and the Eq.~(\ref{I2}) or~(\ref{I3}) can also be analytically solved for $a(t)$:
\begin{equation}
    a(t) = (\frac{3 H_0}{2 n})^{2n/3}\cdot t^{2n/3} .\label{sol1a}
\end{equation}
This solution is also obtained when $I_{1}=0$, which means if we set the unknown conservation quantity $I_1$ as zero, only one final solution (Eqs.~(\ref{sol1}) or~(\ref{sol1a})) can be found according to Noether's theorem in this specific case.

Through another direction, we can obtain the modified Friedman's equation by inserting the Lagrangian~(\ref{Lagrangian f(T)}) into the Einstein's field equation, which read
\begin{eqnarray}
  12 H^2 f_{T} + f  &=& \kappa^2 \rho_{m},  \label{Friedman 1}\\
  48 H^2 f_{TT} \dot{H} - f_{T}(12H^2 + 4\dot{H}) - f &=& p .
\end{eqnarray}
And the Hamiltonian for the from $L\!=\!T(a,\dot{a},T,\dot{T})-V(a,T)$ of the Lagrangian~(\ref{Lagrangian f(T)})
\begin{eqnarray}
  E &=& \dot{T}\frac{\partial L}{\partial \dot{T}} + \dot{a}\frac{\partial L}{\partial \dot{a}} - L \nonumber \\
  &=& T(a,\dot{a},T,\dot{T}) + V(a,T) \nonumber  \\
  &=& -6 f_{T}a \dot{a}^2 + a^3 (f - f_{T}T) \nonumber \\
  &=& a^3(-12H^2 f_{T}- f), \label{E}
\end{eqnarray}
where $T(a,\dot{a},T,\dot{T})\!=\!-6 f_{T} a \dot{a}^2$ defines the kinematic energy term, and $V(a,T)\!=\!a^3 (f - f_{T}T)$ denotes the potential energy term. Combining the first equation (\ref{Friedman 1}) of modified Friedman's equation with Eq.~(\ref{E}), we obtain
\begin{equation}
    E = -a^3 \kappa^2 \rho_{m} = -\kappa^2 \Omega_{m} \rho_{cr,0} = -3H_{0}^2 \Omega_{m},
\end{equation}
where $\rho_{m}\!=\!\rho_{m 0}/ a^3$, $\Omega_{m}\!=\!\rho_{m 0} / \rho_{cr0}$ with $ \rho_{cr0}\!=\!3H_{0}^2 / \kappa^2$ as the critical density and $H_{0}$ is Hubble quantity at present.

Combining the both equations above, we have
\begin{equation}
    3H_{0}^2 \Omega_{m} = a^3(12H^2 f_{T} + f).
\end{equation}
Then taking the form of $f(T)\!=\! s T_{0}^{1-n} T^n$ for dimensional analysis, and $a(0)=1$, the matter density parameter reads
\begin{equation}
    \Omega_{m} = s (2n-1) a^{3} (\frac{T}{T_{0}})^n ,
\end{equation}
which is useful to constrain the parameters $n$ and $s$ from the matter component $\Omega_{m}$.
If $n = 1/2$, $\Omega_{m} = 0$. The universe will be nearly pure dark energy, which is not natural.
\begin{equation}
    s =\frac{\Omega_{m}}{(2n-1)}.
\end{equation}

\subsubsection{Astrophysical data constraints}

\begin{figure}[thbp]
  \begin{center}
  \includegraphics[width=3in]{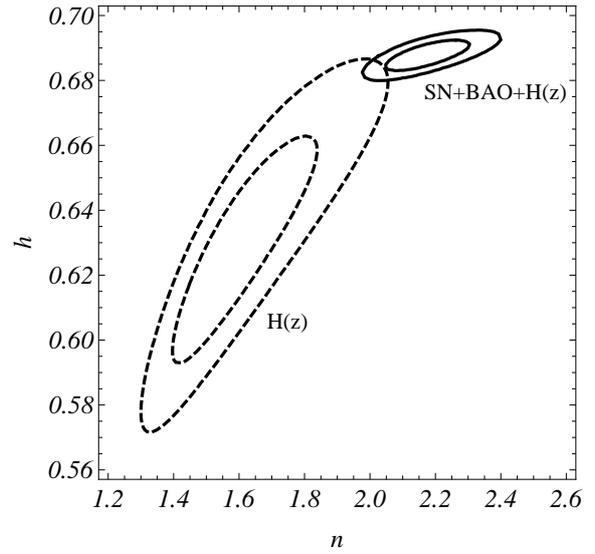}
  \end{center}
  \caption{The 1$\sigma$ and 2$\sigma$ confidence ranges of parameter pair (n,h) (where $h\!=\!H_0/(100~km\cdot s^{-1}\cdot Mpc^{-1})$), constrained by the observational H(z), BAO and SN Ia data-sets. The left contours show the confidence ranges constrained by H(z) data only, while constraints given by the combination of the three data-sets are shown as the contours on the right.}
  \label{fig.chi2}
\end{figure}

Observational Hubble Parameter (OHD or H(z)) data-sets~\cite{OHDdata} , Baryon Acoustic Oscillation (BAO) data-sets (which contains data from 6dF~\cite{6dFdata}, SDSS~\cite{SDSSdata} and Wiggle Z~\cite{WiggleZdata} projects), and the latest Union2.1 SN Ia (SN) data-sets~\cite{SNdata} are taken for model constraining in our work, the best-fit results of parameters are listed in Table.~\ref{table.1} with corresponding reduced minimal $\chi$-square value which is the quantities of the minimum of $\chi$-square divided by the dimension $N$ of parameter-space of model. Besides which the confidence ranges of parameter pair $(n,h)$ are shown in Figs.~\ref{fig.chi2}.

\begin{table}[thbp]
    \caption{Best-fit parameters of $f(T)\sim T^{n}$ model by H(z), SN and BAO with reduced minimal $\chi$-square.}
    \label{table.1}
    \begin{center}
    \begin{tabular}{cccc}
    \hline
    data sets     & reduced $\chi^2_{min}/ N$ & $H_{0}(km/s/Mpc)$ & $n$  \\
    \hline
    H(z)          &20.7488 / 1         &62.7074            &1.58407  \\
    SN            &568.313 / 1         &69.1659            &2.34754  \\
    SN+H(z)       &599.851 / 1         &68.8606            &2.20461  \\
    SN+H(z)+BAO   &603.965 / 1         &68.7674            &2.16819  \\
    \hline
    \end{tabular}
    \end{center}
\end{table}

\begin{figure}[!thbp]
  \begin{center}
  \includegraphics[width=3in]{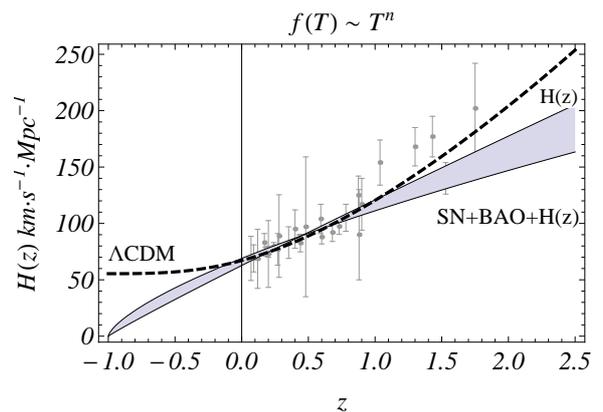}
  \end{center}
  \caption{Contrasts between $\Lambda$CDM and $T^n$ model by the initial conditions $a(0)=1$ and the new $H_{0}$ measurements of planck, where the shadow area is permitted by fitted parameters of various data-sets. For the bigger value of $n$, the H(z) curve becomes more deflected at high red-shift. Inconsistency is drown by the fact that smaller $n$ fits data-sets better at high red-shift, while bigger $n$ performs better at low red-shift.}
  \label{fig.Hz}
\end{figure}

The best-fit results and confidence ranges provided by H(z) alone deviate significantly from the other data-sets, the best-fit parameter $n$ and Hubble quantity $h$ are smaller than those given by SN or BAO data-sets. The deviations are shown in Fig.~\ref{fig.Hz}, from which we find that this $T^n$ model predicts a static universe in the far future with $H(z=-1)\!=\!0$, besides it does not fit CMB observations as well, deviating greatly from the standard $\Lambda$CDM at high red-shift. It is very clear that $f(T)\sim T^{n}$ may not be an ideal approach of $f(T)$ theory also because of its inconsistency as shown in Fig.~\ref{fig.Hz}. So other forms of expressions ie., $f(T)\sim T^m + T^n$, should be taken into consideration.

\subsection{Case \uppercase\expandafter{\romannumeral2}: $f(T)\sim \sum_{n} T^{n}$ and $e^{n T}$}

Since the previous model is not ideal according to astrophysical observation, we shall try other bold approaches which read $f(T)\sim \sum_{n} T^{n}$ and $e^{n T}$. Taking these forms to the determined system of linear PDEs~(\ref{PDEs f(T)}), the solutions of $\tau$, $\phi$, $\psi$ and $B$ for both forms are respectively identical:
\begin{eqnarray}
  \tau &=& C_{1},  \\
  \psi &=& 0, \\
  \phi &=& 0, \\
  B &=& const.
\end{eqnarray}
There exists only one corresponding Noether symmetry generator which reads
\begin{equation}
  X_{1} = \frac{\partial}{\partial t}.
\end{equation}
The first integrals are derived as
\begin{equation}
  I_{1~a} = \sum_{n} \big[(2n-1)a^3 T^n\big], \label{I1a}
\end{equation}
and
\begin{equation}
  I_{1~b} = e^{n T} a^3 (2n T - 1),
\end{equation}
where $I_{1~a}$ is the first integral of the symmetry
of $f(T)\sim \sum_{n} T^{n}$,
and $I_{1~b}$ is the first integral of the symmetry
of $f(T)\sim e^{n T}$.
The exact solution for $f(T)\sim e^{n T}$ is $a\propto e^{\frac{t}{2\sqrt{-3 n}}}$ and $I_{1~b}=0$, but $H\!=\!const$ is apparently not suitable as a cosmological model which must fits observations. From the two cases discussed above we shall draw our conclusion that neither the simple $T^n$ model nor the extremely compact $e^{nT}$ model is acceptable theoretically and practically (by which we mean that any acceptable model should be both theoretically reasonable and observationally permissible). Since the two proposals have failed in deriving appropriate cosmological model discussed above, in the following, we will consider the truncated expressions of case II, like $T^n+T^m$ or $T+T^n$, which remain as another possibility.

\subsection{Case \uppercase\expandafter{\romannumeral3}: $f(T)\!=\!\alpha T+\beta T^n$}

This case can also be regarded as a modification of case I by an addition of a linear term when considering the transition of the large scale of cosmological models to a smaller one, ie., the scale of solar system, for which we can also constraint the parameter $n$ by corresponding conservation quantities. The form ($T+T^n$) has been researched with constraints from the solar system~\cite{fT_conformal,fT_solar}, and we will show its relation with Noether's theorem and cosmological observations.

For $f(T)\!=\! \alpha T + \beta T^{n}$, the only remained symmetry is
the time translational symmetry and the corresponding first integral deduced from Eq.~(\ref{I1a}) as
\begin{equation}
  I =  \alpha a^3 T + \beta (2n-1)a^3 T^n.
\end{equation}
According to the conservation requirement $\frac{\partial I}{\partial t}\!=\!0$, the Hubble quantity must satisfy the following condition
\begin{equation}
  \frac{\beta}{\alpha} (2n-1)(-6H^2)^{n-1}(3H^2+2n \dot{H}) + 3H^2 + 2 \dot{H} =0 ,\label{c3sol}
\end{equation}
which can be numerically solved as a cosmological model, we find the solutions are valid only when
coefficient $n$ is an integral number or zero, ie., $n\!=\!...,2,1,0,-1,-2...$ ($n\!=\!\frac{1}{2}$ is equivalent to $n\!=\!1$ which can verified by Eq.~(\ref{c3sol}), which indicates that $\alpha T+\beta\sqrt{T}$ is equivalent to $\gamma T$ with the same cosmological solution), otherwise the value of the Hubble quantity given by the solution of Eq.~(\ref{c3sol}) is not real.

\begin{figure}[!thbp]
  \begin{center}
  \includegraphics[width=2.8in]{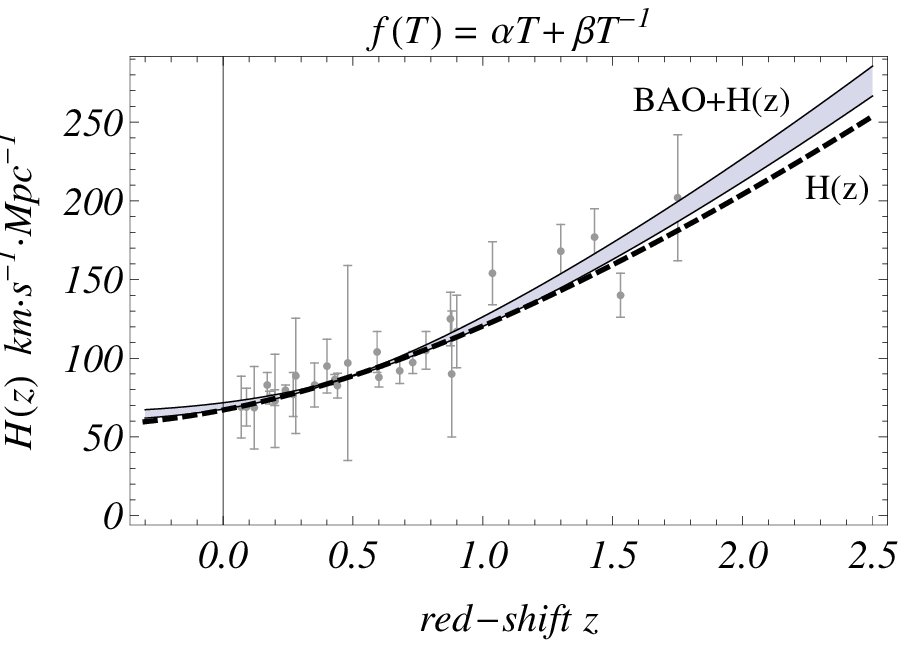}
  \includegraphics[width=2.8in]{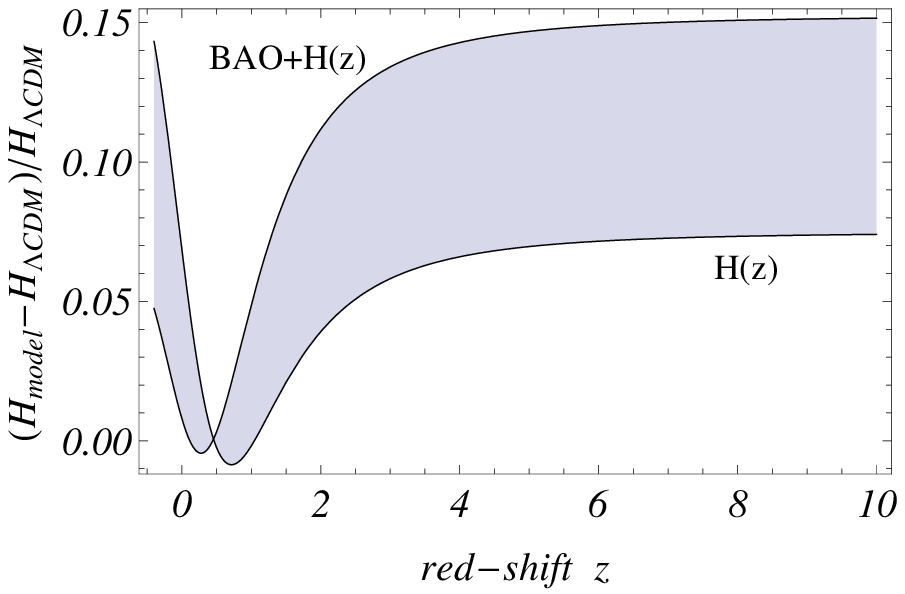}
  \end{center}
  \caption{Contrasts of $H-z$ relation between $\Lambda$CDM (the dashed line) and $f(T)\!=\!\alpha T+\beta T^{-1}$ model with permitted parameter values (the shadow area). Reduction of cosmic age is inevitable since the Hubble quantity depicted by this specific $f(T)$ modification is larger than the standard one as shown by figure on the right.}
  \label{fig.HzaT+bT^-1}
\end{figure}

\begin{figure}[!h]
  \begin{center}
  \includegraphics[width=2.8in]{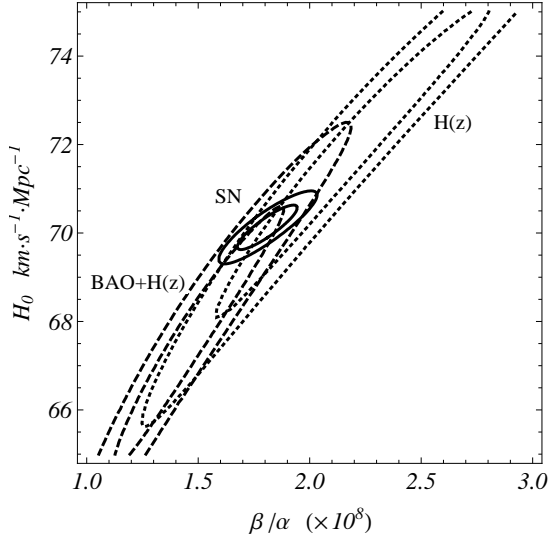}
  \end{center}
  \caption{1\textbf{$\sigma$} and 2 $\sigma$ confidence ranges constrained by various data-sets, where the solid lines represent constraints from SN alone, dashed and dotted lines represent constraints from BAO+H(z) and H(z) respectively. The overlapped region give a tight restriction for the parameter ratio $\beta/\alpha$ between $1.7\times 10^8$ and $1.9\times 10^8$.}
  \label{fig.CRaT+bT^-1}
\end{figure}

Notice that, when the coefficient $n\!=\!0$,
the $\Lambda$CDM is retrieved.
And we find out that $f(T)\!=\!\alpha T+\beta T^{-1}$ is the best model of this case, which can explain and mimic the behavior of dark energy fairly well,
and slightly deviates from the standard model as shown
in the Figs.~\ref{fig.HzaT+bT^-1} for detail.
The magnitude of the order of coefficient ratio $\beta / \alpha$ is as high as $10^8$ constrained by various astrophysical data-sets (the best-fit parameters are shown in Table.~\ref{table.2} with confidence ranges given in Fig.~\ref{fig.CRaT+bT^-1}).

\begin{table}[!h]
    \caption{Best-fit parameters of $f(T)\sim \alpha T+\beta T^{-1}$ model by H(z), SN and BAO.}
    \label{table.2}
    \begin{center}
    \begin{tabular}{cccc}
    \hline
    data sets     & reduced $\chi^2_{min}/N$ & $H_{0}(km/s/Mpc)$ & $ \beta/\alpha (10^{8}) $  \\
    \hline
    H(z)          &14.958 / 2          &71.7287            &2.1564  \\
    SN            &553.11 / 2         &70.1287            &1.8085  \\
    BAO+H(z)      &27.8209 / 2         &67.6546            &1.4702  \\
    SN+H(z)+BAO   &592.494 / 2        &70.061             &1.7930  \\
    \hline
    \end{tabular}
    \end{center}
\end{table}

When the magnitude of torsion $T$ is small on large scales,
the $\beta /T$ term plays a leading role in $f(T)$,
explaining the cosmological effect of dark energy.
When the value of $T$ is much higher on small scales, ie.,
in the solar system, the leading term is
$\alpha T$ which is in accordance with the principle of general relativity.
Besides, the constraining results predict a magnitude of the difference
of $T$ between its small-scale value and large-scale value,
which should be no less than the order of $10^4$.

\section{The distinctions between $\Lambda$CDM and $f(T)$ gravity}

What are the distinctions between $\Lambda$CDM and modified gravity, such as $f(T)$ gravity?
We use the Noether theorem to research this question. According to the distinctions between $\Lambda$CDM and modified gravity, then we can distinguish them from theoretical discussions and the cosmological observations. As we all know, the Teleparallel Equivalent of
General Relativity behaves equivalent to Einstein's theory of general relativity.
So the $f(T)\!=\!-T + const$ is dynamic equivalent to $R + const$, which is easy to be proved.
The $\Lambda$CDM model is the simplest model and accords with the observations.
Thus, we can use $f(T)\!=\!-T + \Lambda$ to describe the cosmology behavior of $\Lambda$CDM.
According to the introduction above, taking the form to the determined system of linear PDEs~(\ref{PDEs f(T)}), the solutions of $\tau$, $\psi$ and $B$ read
\begin{eqnarray}
  \tau &=& \frac{1}{2} C_{2} \frac{e^{\frac{1}{2} \sqrt{6} \sqrt{\Lambda} t}}{\Lambda}
  + \frac{1}{2} C_{1} \frac{e^{-\frac{1}{2} \sqrt{6} \sqrt{\Lambda} t}}{\Lambda} + C_{6}, \\
  \psi &=& \frac{\sqrt{6}}{12\sqrt{\Lambda} a^2} \bigg( - a^{3/2} C_{4} e^{-\frac{1}{4} \sqrt{6} \sqrt{\Lambda} t} + a^{3/2} C_{3} e^{\frac{1}{4} \sqrt{6} \sqrt{\Lambda} t} \nonumber \\
  &+& a^3 C_{2}e^{\frac{1}{2} \sqrt{6} \sqrt{\Lambda} t} - a^3 C_{1} e^{-\frac{1}{2} \sqrt{6} \sqrt{\Lambda} t} \bigg), \\
  B &=& C_{5} + a^{3/2} C_{4} e^{-\frac{1}{4} \sqrt{6} \sqrt{\Lambda} t}
  + a^{3/2} C_{3} e^{\frac{1}{4} \sqrt{6} \sqrt{\Lambda} t} \nonumber \\
  &+& a^3 C_{2} e^{\frac{1}{2} \sqrt{6} \sqrt{\Lambda} t}
  + a^3 C_{1} e^{-\frac{1}{2} \sqrt{6} \sqrt{\Lambda} t}.
\end{eqnarray}
These results are entirely consistent with the ones of $f(R)\!=\!R + \Lambda$, so the following discussions are also applicable to the $\Lambda$CDM model. It is need to note that $\phi$ is arbitrary, and the $C_{i}$ are the normalized orthogonal coefficients. By setting one of the $C_{i}$ to $1$ and the others to $0$, we can get the corresponding Noether symmetry generators, which are deduced as
\begin{eqnarray}
  X_{1} &=& - \frac{\sqrt{6} e^{-\frac{1}{4} \sqrt{6} \sqrt{\Lambda} t}}{12 \sqrt{a}\sqrt{\Lambda}} \frac{\partial}{\partial a} + \phi \frac{\partial}{\partial T} \\
  X_{2} &=& \frac{\sqrt{6} e^{\frac{1}{4} \sqrt{6} \sqrt{\Lambda} t}}{12 \sqrt{a}\sqrt{\Lambda}} \frac{\partial}{\partial a} + \phi \frac{\partial}{\partial T} \\
  X_{3} &=& \frac{e^{\frac{1}{2} \sqrt{6} \sqrt{\Lambda} t}}{2 \Lambda} \frac{\partial}{\partial t} + \frac{\sqrt{6} a e^{\frac{1}{2} \sqrt{6} \sqrt{\Lambda} t}}{12 \sqrt{\Lambda}} \frac{\partial}{\partial a} + \phi \frac{\partial}{\partial T} \\
  X_{4} &=& \frac{e^{-\frac{1}{2} \sqrt{6} \sqrt{\Lambda} t}}{2 \Lambda} \frac{\partial}{\partial t} -  \frac{\sqrt{6} a e^{-\frac{1}{2} \sqrt{6} \sqrt{\Lambda} t}}{12 \sqrt{\Lambda}} \frac{\partial}{\partial a} + \phi \frac{\partial}{\partial T} \\
  X_{5} &=& \phi \frac{\partial}{\partial T} \\
  X_{6} &=& \frac{\partial}{\partial t} + \phi \frac{\partial}{\partial T}
\end{eqnarray}
If we define the $X_{6}$ as the time translational symmetry of energy conservation, so the $\phi$ should be vanished. If not supposed that, then we would lose the time translational symmetry in this system, which does not conform to the fact. So the system reduces to five symmetries.
Moreover, the corresponding first integrals read
\begin{eqnarray}
  I_{1} &=& \bigg(1 + \frac{\sqrt{6}}{\sqrt{\Lambda}} H \bigg) a^{3/2} e^{-\frac{1}{4} \sqrt{6} \sqrt{\Lambda} t}, \\
  I_{2} &=& \bigg(1 - \frac{\sqrt{6}}{\sqrt{\Lambda}} H \bigg) a^{3/2} e^{\frac{1}{4} \sqrt{6} \sqrt{\Lambda} t}, \\
  I_{3} &=& \bigg(\frac{1}{2} + \frac{3}{\Lambda} H^2 - \frac{\sqrt{6}}{\sqrt{\Lambda}} H \bigg) a^3 e^{\frac{1}{2} \sqrt{6} \sqrt{\Lambda} t}, \\
  I_{4} &=& \bigg(\frac{1}{2} + \frac{3}{\Lambda} H^2 + \frac{\sqrt{6}}{\sqrt{\Lambda}} H \bigg) a^3 e^{-\frac{1}{2} \sqrt{6} \sqrt{\Lambda} t}, \\
  I_{6} &=&  a^3 ( 6H^2 - \Lambda),
\end{eqnarray}
among which, we firstly work out that $I_{6}$, $I_{1}$ and $I_{2}$ have the same solution, according to the conservation requirement $\frac{\partial I}{\partial t}\!=\!0$.
\begin{equation}\label{Tconst}
    a(t)^{3/2} = Q_{2} e^{-\frac{1}{4} \sqrt{6} \sqrt{\Lambda} t} - Q_{1} e^{\frac{1}{4} \sqrt{6} \sqrt{\Lambda} t}.
\end{equation}
Moreover, the $I_{3}$ and $I_{4}$ have the same degenerate solution. As we all know, the $\Lambda\!\approx\!H_{0}^2\!=\!(2.1332 h \times 10^{-42} Gev)^2$, $h \!\approx\! 0.7$ and the cosmic age by Planck is about 13.8 billion years, so the $\sqrt{\Lambda} t$ is about $9.75\times 10^{-39} J\cdot s$, which is quite small. Because the university is acceleration, the later term is major term for now and future and the $Q_{1}$ should be a negative number. The scaler factor tends to be
\begin{equation}
    a(t) \propto e^{\frac{\sqrt{\Lambda}}{\sqrt{6}} \cdot t}.
\end{equation}
As a result, the Hubble parameter tends to be constant for the future, because of $H_{f}\!=\!\frac{\sqrt{\Lambda}}{\sqrt{6}}$ (The subscript $f$ represents the future).
We can also expand the solution for the small quantity of $\sqrt{\Lambda} t$ as
\begin{eqnarray}
    a(t)^{3/2} &=& (Q_{2}-Q_{1}) - \frac{1}{4} \sqrt{6} \sqrt{\Lambda} t (Q_{2}+Q_{1}) \nonumber \\
    &+& \frac{3}{16} \Lambda t^2 (Q_{2}-Q_{1}) + \cdots
\end{eqnarray}
Then we can truncate it to the first order, and when $-(Q_{2}+Q_{1}) \gg Q_{2}-Q_{1}$ (this reduces to $Q_{2}\! < \! 0$), so the $a(t)\! \propto \! t^{2/3}$ as a result, which is the solution for the universe with dust as the main ingredient. Thus the coefficients $Q_{1}$ and $Q_{2}$ of scaler factor $a(t)$ are both negative. Moreover, we can calculate the precision form of Hubble parameter as
\begin{equation}
    H = \frac{\sqrt{\Lambda}}{\sqrt{6}} \frac{Q_{1} e^{\frac{1}{4} \sqrt{6} \sqrt{\Lambda} t} + Q_{2} e^{-\frac{1}{4} \sqrt{6} \sqrt{\Lambda} t}}{Q_{1} e^{\frac{1}{4} \sqrt{6} \sqrt{\Lambda} t} - Q_{2} e^{-\frac{1}{4} \sqrt{6} \sqrt{\Lambda} t}}.
\end{equation}

Then we can simplify the first integrals or conserved quantities by the Eq.~(\ref{Tconst}), so we find that
\begin{eqnarray}
  I_{1} &=& 2Q_{1}, \\
  I_{2} &=& 2Q_{2}, \\
  I_{3} &=& \frac{I_{1}^{2}}{2} = 2Q_{1}^2, \\
  I_{4} &=& \frac{I_{2}^{2}}{2} = 2Q_{2}^2, \\
  I_{6} &=& 4Q_{1}Q_{2}\Lambda.
\end{eqnarray}
Due to the negative coefficients $Q_{1}$ and $Q_{2}$, the conserved quantities $I_{1}$ and $I_{2}$ are also negative, which may not be directly determined by observation. But the quantity of energy conservation $I_{6}$ and the additional $I_{3}$ and $I_{4}$ are positive, which is natural for the quantity of energy conservation. Moreover, they mean that the constant $\Lambda$ is relation with some sort of total energy, and there are two other types of observable characteristics of conservation. It's possible to obtain the coefficients $Q_{1}$ and $Q_{2}$ by data fitting from H(z), BAO and SN. It indicates three significant measures.

As we discussed above, the best modified $f(T)$ model is quite different from the $\Lambda$CDM model. The $\Lambda$CDM model has four additional symmetries, two positive and two negative conserved quantities.

\section{Conclusions}
In the modified gravity theories, the explicit expressions of Lagrangian decide the exact solutions of the field equations and are important to gain insight into the physical content of the theories.
While the Noether's theorem offers us a useful tool for researching the conserved quantities and symmetries of modified gravity theories, especially $f(T)$ theory investigated in this article.

As the above research, in the simplest case where $f(T)\sim T^{n}$, the full set of symmetries are expressed corresponding to generators $X_1$, $X_2$ and $X_3$ as shown by Eqs.~(\ref{ftc1x1})-(\ref{ftc1x3}), among which $X_2$ is related to energy conservation or Hamilton condition. $X_1$ is a special additional symmetry which contains $t\frac{\partial}{\partial t}$ term. $X_3$ is an additional symmetry, which may be related to the conformal symmetry. Nevertheless, the cosmological solution of symmetric generators $X_2$ and $X_3$ possess the same form as Eq.~(\ref{sol1}), which can also be derived by setting $I_1\!=\!0$. Although the parameter $n$ can be fitted by astrophysical data, the model $f(T)\sim T^{n}$ is not self-consistent and deviates greatly from the standard depiction of the universe.

We also tried the exponential ($e^{n T}$) and polynomial ($\sum_{n} T^{n}$) forms of $f(T)$ theory, since the exponential ($e^{n T}$) are not theoretically acceptable (predicting a universe with constant Hubble quantity), and the expressions need to be truncated.

Not only as a specific truncated form of case II, it is also meaningful to add $T$ to the simplest form ($f(T)\sim T^n$) of case I to make it much more reasonable and suitable under scale transition, ie. from large scales which could be larger than $100~Mpc$ to the scale of the solar system. We specifically obtained $\alpha T+\beta T^{-1}$ as best fitted model in our present work. And $\alpha T+\beta\sqrt{T}$ is equivalent to $\gamma T$ for the same solution. As shown above, astrophysical data-sets provide pleasing constraints on $\beta /\alpha$ and the Hubble quantity as well, where the magnitude of the order of the coefficients' ratio $\beta /\alpha$ which has seldom been investigated in previous work can be also regarded as a restriction for the magnitude of torsion on different scales. When the torsion is small on large-scales, the $1/T$ term plays the leading role, explaining the phenomenon of dark energy. On the other hand, when $T$ grows much large on small-scales, ie., in the solar system, the leading term is $T$ itself which is equivalent to the general relativity. From the view of phenomenology, the critical magnitude between $\alpha T$ and $\beta T^{-1}$ is similar with the critical velocity of MOND theory~\cite{MOND}. The cosmic age predicted by this specific model is reduced by a tiny amount, by introducing higher order $\gamma T^m$ term to the form of $\alpha T+\beta T^{-1}$ can alleviate that drawback.

The time translational symmetry derived from Noether's theorem is mainly concerned in this work. Noether's theorem is also a feasible approach to discuss the conformal symmetry, which is used to discuss the Dark Energy by many modified gravity theories, ie., the Jordan frame and Einstein frame are connected by the conformal transformation.

The methodology presented in this work can also be adopted for $f(R)$ theory. Similarly, $f(R)\sim R^n$ can not satisfy the early-stage and the late-stage cosmic acceleration at the same time, which has been pointed by other related work~\cite{fR_inconsistent}. Concerning a modified expression which resembles the work done in this article is inevitable.

More important is that we find the significant differences between $\Lambda$CDM and $f(T)$ gravity. The $\Lambda$CDM model has four additional symmetries and corresponding conserved quantities, but the best modified $f(T)$ model has only time translational symmetry. The constant $\Lambda$ is relation with some sort of total energy, and there are two other types of observable positive characteristics of conservation. If we can confirm the number of symmetries or conserved quantities by observations, we will be able to judge which theory is more appropriate or right.

\section*{Acknowledgements}

This work is partly supported by National Natural
Science Foundation of China under Grant Nos. 11075078 and 10675062
and by the project of knowledge Innovation Program (PKIP) of Chinese
Academy of Sciences (CAS) under the grant No. KJCX2.YW.W10 through
the KITPC where we have initiated this present work.
Han Dong would also like to thank his parents for the support to
complete master's graduate studies.

\end{document}